%% file: main.tex
\documentclass[sigplan,nonacm]{acmart}

\usepackage{tikz}

\usepackage{caption}

\usepackage{booktabs}
\usepackage{enumitem}
\usepackage{balance}
\usepackage{graphicx}
\usepackage{lipsum}  
\usepackage{url}
\usepackage{color}
\usepackage{listings}
\usepackage{xspace}
\usepackage{makecell}
\usepackage{fancyvrb}
\usepackage{verbatim}
\usepackage{etoolbox}
\usepackage{multirow}
\usepackage{comment}
\usepackage{hyperref}
\usepackage[normalem]{ulem}
\usepackage[noend]{algpseudocode}
\usepackage{algorithm}
\usepackage[normalem]{ulem}
\usepackage{float}
\usepackage{subcaption}

\input{defs}

\setlist{nosep}
\newtheorem{theorem}{Theorem}

\makeatletter
\def\UrlAlphabet{%
      \do\a\do\b\do\c\do\d\do\e\do\f\do\g\do\h\do\i\do\j%
      \do\k\do\l\do\m\do\n\do\o\do\p\do\q\do\r\do\s\do\t%
      \do\u\do\v\do\w\do\x\do\y\do\z\do\A\do\B\do\C\do\D%
      \do\E\do\F\do\G\do\H\do\I\do\J\do\K\do\L\do\M\do\N%
      \do\O\do\P\do\Q\do\R\do\S\do\T\do\U\do\V\do\W\do\X%
      \do\Y\do\Z}
\def\UrlDigits{\do\1\do\2\do\3\do\4\do\5\do\6\do\7\do\8\do\9\do\0}
\g@addto@macro{\UrlBreaks}{\UrlOrds}
\g@addto@macro{\UrlBreaks}{\UrlAlphabet}
\g@addto@macro{\UrlBreaks}{\UrlDigits}
\makeatother

\copyrightyear{2024}
\acmYear{2024}
\setcopyright{acmlicensed}\acmConference[ASPLOS '24]{29th ACM International Conference on Architectural Support for Programming Languages and Operating Systems, Volume 3}{April 27-May 1, 2024}{La Jolla, CA, USA}
\acmBooktitle{29th ACM International Conference on Architectural Support for Programming Languages and Operating Systems, Volume 3 (ASPLOS '24), April 27-May 1, 2024, La Jolla, CA, USA}
\acmDOI{10.1145/3620666.3651383}
\acmISBN{979-8-4007-0386-7/24/04}
\begin{document}

\title{Optimal Kernel Orchestration for Tensor Programs with \sys}

\author{Muyan Hu}\authornote{Work started as part of an internship at CMU.} 
    \affiliation{\institution{University of Illinois at Urbana-Champaign}\city{Urbana-Champaign}\state{IL}\country{USA}}
    \orcid{https://orcid.org/0009-0001-4096-0511}
    \email{muyanhu2@illinois.edu}
\author{Ashwin Venkatram}\authornote{Equal contribution.}
    \affiliation{\institution{Advanced Micro Devices}\city{San Jose}\state{CA}\country{USA}}
    \orcid{https://orcid.org/0009-0005-4661-0060}
    \email{ashwinve@alumni.cmu.edu}
\author{Shreyashri Biswas}\authornotemark[2]
    \affiliation{\institution{Carnegie Mellon University}\city{Pittsburgh}\state{PA}\country{USA}}
    \orcid{https://orcid.org/0009-0002-6656-1030}
    \email{sbiswas2@andrew.cmu.edu}
\author{Balamurugan Marimuthu}\authornotemark[2]
    \affiliation{\institution{Sambanova Systems}\city{Palo Alto}\state{CA}\country{USA}}
    \orcid{https://orcid.org/0000-0001-6292-5066}
    \email{bmarimut@alumni.cmu.edu}
\author{Bohan Hou}
    \affiliation{\institution{Carnegie Mellon University}\city{Pittsburgh}\state{PA}\country{USA}}
    \orcid{https://orcid.org/0000-0001-5718-3387}
    \email{bohanhou@andrew.cmu.edu}
\author{Gabriele Oliaro}
    \affiliation{\institution{Carnegie Mellon University}\city{Pittsburgh}\state{PA}\country{USA}}
    \orcid{https://orcid.org/0000-0001-5406-0736}
    \email{goliaro@andrew.cmu.edu}
\author{Haojie Wang}
    \affiliation{\institution{Tsinghua University}\city{Beijing}\country{China}}
    \orcid{https://orcid.org/0000-0003-4605-148X}
    \email{wanghaojie@tsinghua.edu.cn}
\author{Liyan Zheng}
    \affiliation{\institution{Tsinghua University}\city{Beijing}\country{China}}
    \orcid{https://orcid.org/0000-0001-7327-748X}
    \email{zhengly20@mails.tsinghua.edu.cn}
\author{Xupeng Miao}
    \affiliation{\institution{Carnegie Mellon University}\city{Pittsburgh}\state{PA}\country{USA}}
    \orcid{https://orcid.org/0000-0002-9371-8358}
    \email{xupeng@cmu.edu}
\author{Jidong Zhai}
    \affiliation{\institution{Tsinghua University}\city{Beijing}\country{China}}
    \orcid{https://orcid.org/0000-0002-7656-6428}
    \email{zhaijidong@tsinghua.edu.cn}
\author{Zhihao Jia}
    \affiliation{\institution{Carnegie Mellon University}\city{Pittsburgh}\state{PA}\country{USA}}
    \orcid{0000-0002-1270-5185}
    \email{zhihao@cmu.edu}

\begin{abstract}
\input{abstract}

\end{abstract}
\renewcommand{\shortauthors}{Hu et al.}


\maketitle

\input{introduction}

\input{overview}
\input{fission}

\input{ilp}
\input{implementation}
\input{evaluation}
\input{related}
\input{future_work}
\input{conclusion}
\input{appendix}

\balance

\input{ref.bbl}
\bibliographystyle{plain}

\end{document}

%% file: defs.tex
\usepackage{xcolor}
\usepackage{amsmath}
\usepackage{amsthm}
\usepackage{thmtools}
\usepackage{nameref,cleveref}

\definecolor{darkgreen}{rgb}{0,0.5,0}

\definecolor{purple}{rgb}{0.3,0.0,0.6}

\definecolor{auburn}{rgb}{0.43,0.21,0.1}

\definecolor{gray}{rgb}{0.7,0.7,0.7}

\definecolor{dgray}{rgb}{0.5,0.5,0.5}

\definecolor{darkgreen}{rgb}{0,0.5,0}

\newcommand{\er}[1]{\mbox{\rm\em #1}}
\newcommand{\m}[1]{\mathcal{#1}}

\newcommand{\sys}{Korch\xspace}

\newtheorem{definition}{Definition}

\crefformat{section}{\S#2#1#3}
\crefformat{subsection}{\S#2#1#3}
\crefrangeformat{section}{\S#2#1#3}
\crefrangeformat{subsection}{\S#2#1#3}

%% file: abstract.tex
{\em Kernel orchestration} is the task of mapping the computation defined in different operators of a deep neural network (DNN) to the execution of GPU kernels on modern hardware platforms.
Prior approaches optimize kernel orchestration by greedily applying {\em operator fusion}, which fuses the computation of multiple operators into a single kernel, and miss a variety of optimization opportunities in kernel orchestration.

This paper presents \sys, a tensor program optimizer that discovers {\em optimal} kernel orchestration strategies for tensor programs. Instead of directly fusing operators, \sys first applies {\em operator fission} to decompose tensor operators into a small set of basic tensor algebra primitives.
This decomposition enables a diversity of fine-grained, inter-operator optimizations.
Next, \sys optimizes kernel orchestration by formalizing it as a constrained optimization problem, leveraging an off-the-shelf binary linear programming solver to discover an optimal orchestration strategy, and generating an executable that can be directly deployed on modern GPU platforms.
Evaluation on a variety of DNNs shows that \sys outperforms existing tensor program optimizers by up to 1.7$\times$ on V100 GPUs and up to 1.6$\times$ on A100 GPUs.
\sys is publicly available at \url{https://github.com/humuyan/Korch}.

%% file: introduction.tex
\section{Introduction}
\label{sec:intro}

Enabling high-performance execution of deep neural networks (DNNs) is critical in many ML applications, including autonomous driving~\cite{heisley1991autodriving}, object detection~\cite{papageorgiou2000trainable}, machine translation~\cite{bert}, and content generation~\cite{song2020score}. A key bottleneck is the mapping of a DNN's high-level representation to the corresponding hardware-specific computation that can be dispatched to accelerators. 
A DNN model is represented as a directed acyclic graph (DAG), where each node is a tensor operator (e.g., matrix multiplication or convolution) and each edge represents a tensor (i.e., an $n$-dimensional array) shared between operators. On the other hand, computation on modern hardware accelerators (e.g., GPUs and TPUs) is organized in {\em kernels}, each of which is a function simultaneously executed by multiple hardware threads in a single-program-multiple-data (SPMD) fashion~\cite{darema1988spmd}.

To map operators defined in a DNN model to kernels, existing deep learning frameworks~\cite{tvm, abadi2016tensorflow, tensorrt, pytorch} apply {\em operator fusion}, a common form of optimization that fuses the computation of multiple tensor operators into a single kernel to reduce kernel launch overhead and eliminate unnecessary instantiation of intermediate results.
For example, TensorFlow~\cite{abadi2016tensorflow}, TensorRT~\cite{tensorrt}, and TVM~\cite{tvm} consider operator fusion rules manually designed by domain experts and greedily apply these fusion rules whenever possible.
DNNFusion~\cite{dnnfusion} classifies operators into five types based on the data dependencies between the input and output elements of each operator, which allows DNNFusion to develop rules for specific combinations of types instead of combinations of operators. Although existing rule-based operator fusion approaches improve the performance of DNNs on modern accelerators, they have two key limitations.

First, kernel fusion at the operator level is too coarse-grained to unlock all potential optimizations, as some operators achieve suboptimal performance when running in a single kernel.
A potential class of optimizations is to decompose these operators into fine-grained components, and fuse components with identical parallelism degrees and memory access patterns across operators into a single kernel.
However, these optimizations are excluded in existing frameworks since they cannot be directly represented as operator fusions.
One such example is the commonly-used {\tt softmax} operator, which converts a vector of numbers to a vector of probabilities:
\begin{equation}
    \label{eqn:softmax}
    {\tt softmax}(x_i) = {e^{x_i}} / {\sum_j e^{x_j}}
\end{equation}
A {\tt softmax} includes element-wise computation (i.e., the exponential function), vector-wise aggregation (i.e., the summation term in the denominator), and vector-wise broadcast (i.e., the scaling factor $s = {1}/{\sum_j{e^{x_j}}}$ is used to compute all output elements). The three components in {\tt softmax} involve different degrees of parallelism and memory access patterns, making it challenging to generate a single high-performance kernel for {\tt softmax}. Meanwhile, generating a dedicated kernel for each of these components in {\tt softmax} is also suboptimal since this would result in high kernel launch overhead and unnecessary instantiation of intermediate results. 
We could obtain significant performance improvement by performing inter-operator fusion of the operator's subcomponents at a finer-grained level (e.g., fusing the element-wise computation of {\tt softmax} with element-wise components in the previous or next operator).

A second issue with existing operator fusion approaches is their reliance on manually-designed rules to fuse operators greedily, which require significant engineering effort and may miss subtle optimizations that are hard to discover manually. Prior work~\cite{jia2019taso} has also shown that operation fusion rules designed for one type of device do not directly apply to other device types (e.g., a different generation of GPUs), which is problematic as an increasingly diverse set of AI accelerators have been introduced to deploy DNN computation.

In this paper, we first identify {\em kernel orchestration}, the task of orchestrating the computation defined in different operators of a DNN model to the execution of kernels on modern hardware accelerators.
This task has been mostly ignored by prior approaches, resulting in missed optimization opportunities. To this end, we introduce \sys, an automated and systematic approach to discovering {\em optimal} kernel orchestration strategies for tensor programs. Different from existing deep learning frameworks, which use rule-based strategies to fuse operators into kernels, \sys first applies {\em operator fission} to decompose each tensor operator into a small set of basic tensor algebra {\em primitives}. 
Computation defined in each tensor primitive involves the same parallelism degree and data access pattern, making it efficient to execute a tensor primitive in a single kernel. Therefore, primitive graphs provide a more flexible representation for organizing computation of a DNN model compared to the commonly used computation graphs.

Second, to map a primitive graph to kernels, \sys uses a depth-first search algorithm to identify all possible kernels for a primitive graph. \sys formulates kernel orchestration as a constrained optimization problem and uses an off-the-shelf binary linear programming solver to discover an {\em optimal} strategy to map tensor algebra primitives into kernels while minimizing execution cost.
Finally, \sys uses the strategy identified by the solver to construct kernels and generate an executable that can be directly deployed on modern GPU platforms. 

We evaluate \sys on both standard DNN benchmarks as well as emerging workloads and compare \sys with existing tensor program optimizers. Experimental results show that \sys outperforms existing frameworks by up to 1.7$\times$ and 1.6$\times$ on NVIDIA V100 and A100 GPUs, respectively. 
For most DNNs in our evaluation, the best kernel execution strategy discovered by \sys is outside of the search space of existing operator fusion strategies.

This paper makes the following contributions:
\begin{itemize}
    \item We identify the kernel orchestration task and unlock a variety of previously missing kernel optimizations.
    \item We introduce operator fission to decompose operators into fine-grained primitives to enable cross-operator fusion for maximizing kernel efficiency.
    \item We develop a binary linear programming algorithm to optimally map primitive graphs to kernels.
    \item We implement \sys and thoroughly evaluate it, showing up to 1.7$\times$ improvement over existing approaches.
\end{itemize}

%% file: overview.tex
\section{Overview}
\label{sec:overview}

\begin{figure}
    \centering
    \includegraphics[width=\linewidth]{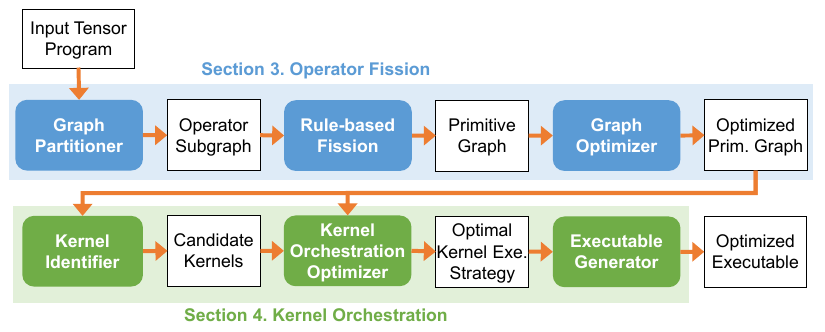}
    \caption{An overview of \sys.}
    \label{fig:overview}
\end{figure}

\Cref{fig:overview} shows an overview of \sys. The input to \sys is a tensor program to be optimized, which is represented as a {\em computation graph} whose nodes and edges are tensor algebra operators and data dependencies between them, respectively. Similar to prior work~\cite{wang2021pet, MetaFlow}, \sys first partitions an input computation graph into smaller subgraphs to reduce the optimization space associated with each subgraph while preserving optimization opportunities.

For each subgraph, \sys's {\em operator fission engine} uses a rule-based approach to decoupling each operator into one or multiple tensor algebra primitives and generating a {\em primitive graph}, which is functionally equivalent to the original computation graph.
\sys's {\em primitive graph optimizer} adopts the superoptimization techniques introduced in prior work~\cite{jia2019taso} and optimizes a primitive graph by using the graph transformations discovered by TASO~\cite{jia2019taso}.
By decoupling tensor operators into fine-grained primitives, \sys enables additional primitive-graph-level optimizations that are infeasible to be performed at the operator graph level (see \Cref{sec:fission}).
This decomposition also allows \sys to perform more flexible kernel orchestration by directly choosing subgraphs of the primitive graph as candidate kernels (see \Cref{sec:mapping}). %

To map tensor algebra primitives to GPU kernels, \sys's {\em kernel identifier} uses a depth-first search algorithm to identify {\em all} possible kernels for a given primitive graph. %
\sys's kernel profiler then measures the runtime performance of each candidate kernel.
\sys formulates kernel orchestration as a {\em binary linear programming} (BLP) problem and uses an off-the-shelf BLP solver to discover an optimal kernel orchestration strategy.
In particular, \sys's {\em kernel orchestration optimizer} takes all identified kernels and their execution latencies as input and employs a BLP solver to select an {\em optimal} strategy, which executes a subset of the candidate kernels and computes the output of the tensor program based on the inputs while minimizing total execution cost.
Finally, the optimal execution strategy is used by \sys's {\em executable generator} to produce an executable for the input tensor program. We describe the operator fission and kernel orchestration components in \Cref{sec:fission} and \Cref{sec:mapping} respectively.

%% file: fission.tex
\section{Operator Fission}
\label{sec:fission}
\begin{figure*}
    \centering
    \subfloat[Comp. graph.]{
        \includegraphics[scale=0.62]{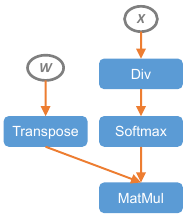}
        \label{fig:comp_graph}
    }
    \subfloat[Primitive graph and transformations.]{
        \includegraphics[scale=0.64]{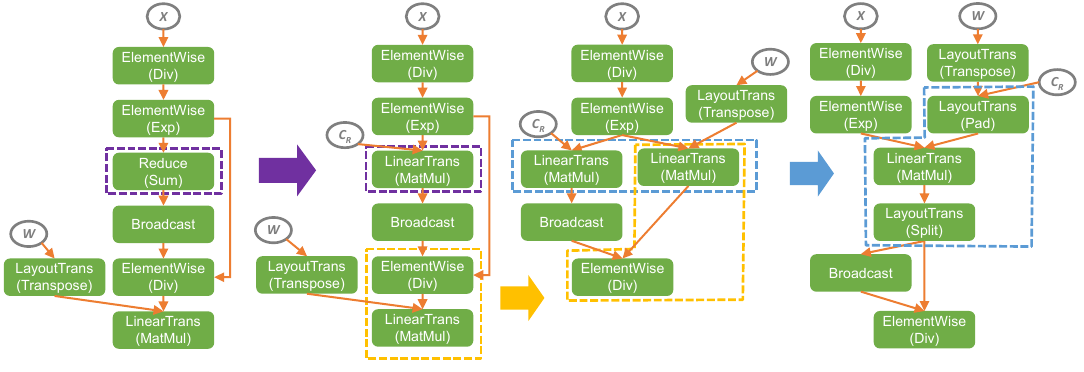}
        \label{fig:primitive_graph}
    }
    \subfloat[Kernel orchestration.]{
        \includegraphics[scale=0.64]{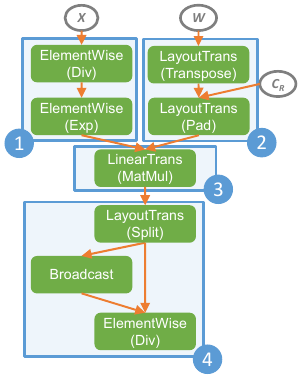}
        \label{fig:kernel_mapping}
    }
    \caption{Operator fission enables subsequent optimizing transformations on primitive graphs. In \Cref{fig:primitive_graph}, the dotted boxes in the same color indicate a transformation on the primitive graph. The combination of the three transformations fuses the reduce primitive in {\tt Softmax} and the subsequent {\tt MatMul} into a single {\tt MatMul}.}
    \label{fig:fission}
\end{figure*}

Existing deep learning frameworks optimize kernel orchestration by applying {\em operator fusion} to opportunistically fuse multiple tensor operators into a single GPU kernel~\cite{tvm, tvm_auto_tuner}.
However, directly applying fusion at the operator level misses a variety of fine-grained, inter-operator optimizations. This is because tensor operators are designed based on their algebraic semantics and properties instead of their implementations on GPUs. As a result, for operators with sophisticated mathematical semantics such as normalization and aggregation, performing all computation defined in such an operator within one kernel results in suboptimal performance.

Based on the above observation, \sys first applies {\em operator fission} to decouple each tensor operator into a small set of tensor algebra {\em primitives}, each of which carries basic algebraic calculation that involves a unified degree of parallelism and data access pattern.
The primitives considered by \sys are divided into four categories based on the relationship between the input and output elements of each primitive.

For each primitive $p$, let $O_p(I)$ denote the output tensor\footnote{Without loss of generality, we assume each primitive has only one output tensor, and our analysis can be directly generalized to primitives with multiple output tensors by analyzing them sequentially.} of running $p$ on $n$ input tensors $I=(I_1,...,I_n)$. For an $m$-dimensional tensor $O$, let $O[\vec{x}]$ denote the output value at position $\vec{x}=(x_1,...,x_m)$. Similarly, let $I_k[\vec{x}]$ denote the input value at position $\vec{x}$ of the $k$-th input tensor $I_k$. We use the above notations to introduce the four categories of tensor algebra primitives considered by \sys.

\paragraph{Elementwise primitives.} For an elementwise primitive $p$, the output tensor shares the same shape and data layout as all input tensors, and the computation for each output element depends on the input elements at the same position:
\begin{equation*}
   O[\vec{x}] = f(I_1[\vec{x}], I_2[\vec{x}], ..., I_n[\vec{x}])
\end{equation*}
Elementwise primitives do not involve layout transformations and can generally be fused with other primitives as a pre- or post-processing step.

\paragraph{Reduce and broadcast primitives.} A reduce primitive takes as input a single tensor and calculates the aggregated result of each row of the input tensor along a given dimension: 
\begin{eqnarray*}
& O[x_1,...,x_{k-1}, x_{k+1},..., x_m]\\
& \quad\quad\quad\quad\quad\quad= \oplus_{x_k} I_1[x_1,...,x_{k-1}, x_k, x_{k+1},...,x_m],
\end{eqnarray*}
which aggregates along the $k$-th dimension of $I_1$, and $\oplus$ is the aggregator. A {\em broadcast} primitive performs the reverse, taking a single input tensor and replicating the input tensor along a given dimension:
\begin{eqnarray*}
& O[x_1,...,x_{k-1}, x_k, x_{k+1},..., x_m]\\
& \quad\quad\quad\quad\quad\quad= I_1[x_1,...,x_{k-1}, x_{k+1},...,x_m].
\end{eqnarray*}

\paragraph{Layout transformation primitives.} A {\em layout transformation} primitive takes a single tensor as input and transforms its data layout without performing arithmetic operations:
\begin{equation*}
O[\vec{x}] = I_1[L(\vec{x})],
\end{equation*}
where $\vec{u} = L(\vec{x})$ is a one-to-one mapping from position $\vec{x}$ of the output tensor to position $\vec{u}$ of the input tensor. 
Note that a layout transformation primitive can have multiple input/output tensors. For example, concatenation and split are considered layout transformation primitives, since they require changing the underlying layouts of the input tensors.

\paragraph{Linear transformation primitives.} Finally, \sys uses {\em linear transformation} primitives to capture the compute-intensive operators in DNNs such as matrix multiplication and convolution. Specifically, a primitive $p$ is considered as a linear transformation primitive if its output is {\em linear} to all input tensors:
\begin{eqnarray*}
& \forall Y, Z, \vec{x}, 1\leq k\leq n. \quad O(I_1,...,I_{k-1}, Y + Z, ...)[\vec{x}] \\
& = O(I_1,...,I_{k-1}, Y, ...)[\vec{x}] + O(I_1,...,I_{k-1}, Z, ...)[\vec{x}],\\
& \alpha\cdot O(I_1,...,I_{k-1}, Y, ...)[\vec{x}] = O(I_1,...,I_{k-1}, \alpha \cdot Y, ...)[\vec{x}]
\end{eqnarray*}

\begin{figure}
    \centering
    \includegraphics[scale=0.42]{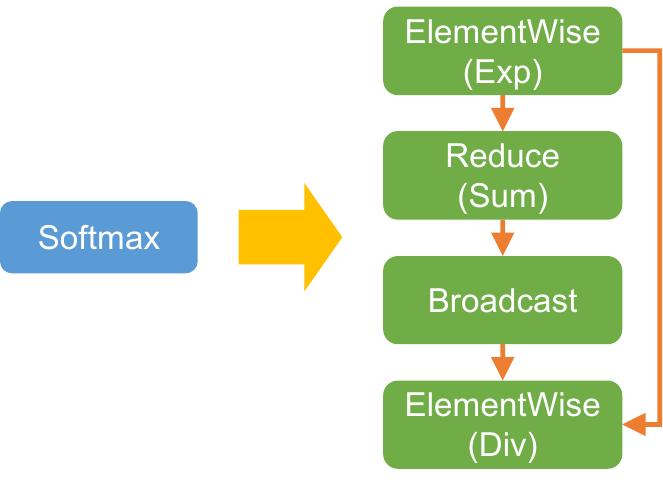}
    \caption{The operator fission rule for {\tt Softmax}.}
    \label{fig:operator_fission}
\end{figure}

\begin{table}[t]
\caption{Examples of some frequently used primitves.}
\label{tab:primitive-example}
\resizebox{\linewidth}{!}{
\centering
\begin{tabular}{c|c}
\toprule
Primitive Type & Representative Operators \\
\midrule
Elementwise & Add, Sub, Mul, Div, Relu, Sqrt, Erf \\
Reduce and broadcast & ReduceSum, ReduceMean, MaxPool, Broadcast \\
Layout transformation & Transpose, Split, Concat, Slice, Pad, Reshape \\
Linear transformation & Conv, GEMM, Batched GEMM \\
\bottomrule
\end{tabular}
}
\end{table}

\Cref{tab:primitive-example} shows representative operators for the four types of primitives.
For each DNN operator, \sys requires developers to specify an operator fission rule, which is used to decouple the DNN operator into the specified primitives.
Each operator fission rule is a transformation from a DNN operator to a functionally equivalent primitive graph. To demonstrate the complexity of designing an operator fission rule, \Cref{fig:operator_fission} shows the operator fission rule for {\tt softmax} (see \Cref{eqn:softmax}), 
which is decomposed into an elementwise exponential, a reduce, a broadcast, and an elementwise division primitive.

Operator fission enables two important classes of optimizations missing in existing deep learning frameworks. First, by decoupling operators into fine-grained primitives, operator fission enables optimizing transformations on primitive graphs; these transformations are impossible to apply at the operator level. \Cref{fig:primitive_graph} shows such an example to optimize multi-head attention in Transformer~\cite{vaswani2017attention}.
First, the reduce primitive of {\tt Softmax} is substituted by a linear transformation primitive (i.e., {\tt MatMul}) by introducing a constant input tensor $C_s$ whose elements are all ones.
Second, the elementwise division primitive of {\tt Softmax} is swapped with the subsequent {\tt MatMul} by applying a transformation discovered by prior work~\cite{jia2019taso}.
Finally, the two {\tt MatMul}s with a shared input are fused by introducing a {\tt Pad}\footnote{The reduce primitive {\tt Sum} can be transformed to a linear transformation, i.e., {\tt MatMul} with a ones vector. Therefore, we should concatenate the ones vector with the other input, which is done by padding ones with it.} and a {\tt Split} primitive.
These transformations allow \sys to fuse the reduce primitive in {\tt Softmax} with a {\tt MatMul}.

A second advantage of operator fission is enabling more flexible kernel orchastration.
Specifically, for the optimized primitive graph in \Cref{fig:kernel_mapping}, \sys maps {\tt Div} and {\tt Exp} into kernel \textcircled{1}, which achieves the same performance as just performing {\tt Div} in a single kernel due to the low arithmetic intensity of elementwise computation.
Similarly, {\tt Transpose} and {\tt Pad} are mapped to kernel \textcircled{2}, which has the same performance as a single {\tt Transpose} kernel.
{\tt MatMul} itself is mapped to kernel \textcircled{3}, since it is compute-intensive and mapping it as a single kernel allows \sys to use kernels provided by CUDA libraries (see details in \Cref{subsec:profiler}).
Finally, the {\tt Split}, {\tt Broadcast}, and {\tt Div} primitives are all mapped to kernel \textcircled{4}.
Both {\tt Split} and {\tt Broadcast} do not introduce arithmetic operations and are cheap to execute.
The performance improvement of this optimization is due to the fact that kernel \textcircled{4} is much faster than launching a {\tt Softmax} kernel.

\paragraph{Supporting new operators.} The four categories of primitives listed above are sufficient to represent a variety of tensor operators used in today's DNNs, including all DNN benchmarks in our evaluation. However, there are tensor operators that cannot be represented using the above primitives. One such operator is {\tt Topk}, which returns the $k$ largest elements of an input tensor along a given dimension. For such operators, \sys treats them as opaque primitives and will only optimize the rest of the primitive graph in graph optimizations.

%% file: ilp.tex
\section{Kernel Orchestration}
\label{sec:mapping}
\begin{figure*}
    \centering
    \subfloat[Primitive graph.]{
        \includegraphics[scale=1]{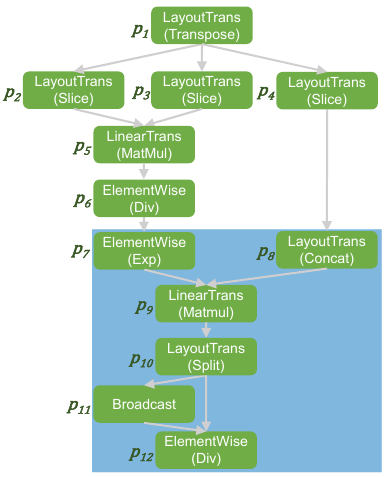}
        \label{fig:primitive_graph_kernel_mapping}
    }
    \hspace{1em}
    \subfloat[Identified kernels.]{
        \includegraphics[scale=0.45]{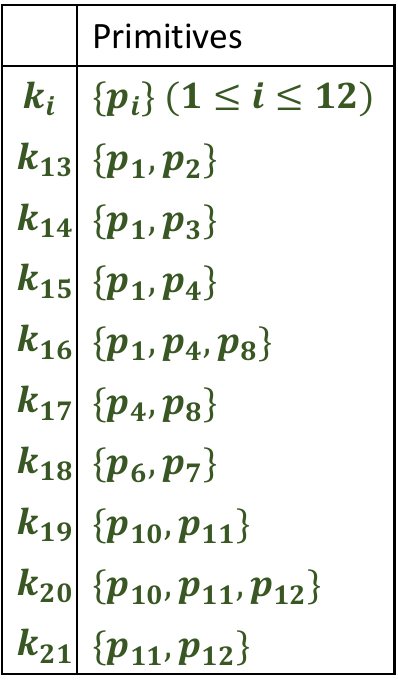}
        \label{fig:kernel_table}
    }
    \subfloat[An optimal kernel orchestration strategy.]{
        \includegraphics[scale=1]{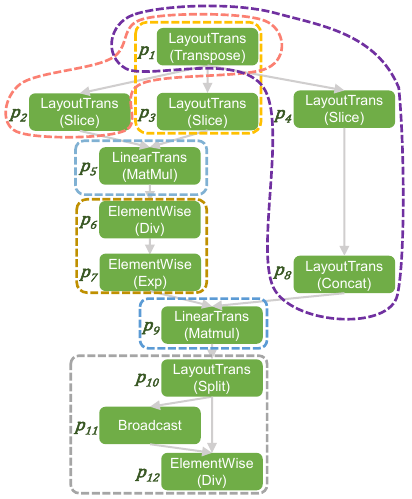}
        \label{fig:kernel_mapping_strategy}
    }
    \caption{An example of kernel orchestration for a subgraph of self attention \cite{vaswani2017attention}. The shadowed subgraph in \Cref{fig:primitive_graph_kernel_mapping} is similar with the primitive graph transformation result in \Cref{fig:fission}.}
\end{figure*}

This section describes \sys's kernel orchestration engine, which takes a primitive graph as an input and discovers an {\em optimal} strategy to map the primitives to kernels while minimizing execution latency. 
A primitive graph is represented as a directed acyclic graph $\m{G}=(\m{P}, \m{E})$, where each node $p\in \m{P}$ is a tensor algebra primitive, and each edge $(p_1, p_2) \in \m{E}$ denotes a tensor that is an output of $p_1$ and an input of $p_2$.
Computations on GPUs are organized as {\em kernels}, each of which is a function simultaneously executed by multiple threads in a single-program-multiple-data (SPMD) fashion.
Each kernel executes one or multiple primitives.
To avoid circular dependencies between kernels (i.e., two kernels depend on each other to start), the primitives included in a kernel must form a {\em convex subgraph} of $\m{G}$.

\begin{definition} [Convex subgraph]
For a primitive graph $\m{G} = (\m{P}, \m{E})$, a set of nodes $\m{P}' \subseteq \m{P}$ forms a convex subgraph of $\m{G}$ if and only if there do {\bf not} exist nodes $p_1, p_2 \in \m{P}'$ and another node $q\in \m{P} \setminus \m{P}'$ such that $p_1 \overset{\m{G}}\leadsto q$ and $q \overset{\m{G}}\leadsto p_2$, where $x \overset{\m{G}}\leadsto y$ indicates that there exists a path from $x$ to $y$ in $\m{G}$.
\end{definition}

For the primitive graph in \Cref{fig:primitive_graph_kernel_mapping}, node set $\{p_1, p_4, p_8\}$ forms a convex subgraph of $\m{G}$ and thus can be executed in a single kernel. On the other hand, we cannot execute node set $\{p_1, p_2, p_5\}$ (a non-convex subgraph of $\m{G}$) in a single kernel, since it has a circular dependency with the kernel that executes $p_3$ (i.e., $p_5$ depends on $p_3$ which depends on $p_1$). \sys uses {\em execution states} to identify all convex subgraphs of a given primitive graph.

\begin{definition} [Execution state]
A set of nodes $\m{P}' \subseteq \m{P}$ is an execution state of a primitive graph $\m{G}=(\m{P},\m{E})$ if for any edge $(p_1, p_2)\in \m{E}$, $p_2 \in \m{P}'$ implies $p_1 \in \m{P}'$.
\end{definition}

Transiting from an execution state to another allows \sys to identify all possible convex subgraphs, as shown in \Cref{thm1}.
The computation of each convex subgraph of $\m{G}$ can be performed in a {\em candidate kernel}.

\begin{theorem}
\label{thm1}
A set of nodes $\m{P}' \subseteq \m{P}$ forms a convex subgraph of $\m{G} = (\m{P}, \m{E})$ if and only if $\m{P}'$ is the difference of two execution states $\m{P}_1$ and $\m{P}_2$: $\m{P}' = \m{P}_1 \setminus \m{P}_2$.
\end{theorem}

\begin{proof}
If $\m{P}'=\m{P}_1\setminus\m{P}_2$, the following will show that $\m{P}'$ is a convex subgraph.
Suppose there exist $p_1,p_2\in\m{P}'$ and $q\in\m{P}\setminus\m{P}'$ such that $p_1 \overset{\m{G}}\leadsto q$ and $q \overset{\m{G}}\leadsto p_2$.
Therefore, $p_1,p_2\in\m{P}_1$ and $p_1,p_2\notin\m{P}_2$. Also we can get $q\notin\m{P}_1$ or $q\in\m{P}_2$.
If $q\notin\m{P}_1$, $q \overset{\m{G}}\leadsto p_2$ and $p_2\in\m{P}_1$ will contradict that $\m{P}_1$ is an execution state.
If $q\in\m{P}_2$, $p_1 \overset{\m{G}}\leadsto q$ and $p_1\notin\m{P}_2$ will contradict that $\m{P}_2$ is an execution state.
In conclusion, the supposition is wrong, so $\m{P}'$ is a convex subgraph.

If $\m{P}'$ is convex graph, let $\m{S}=\{p\mid p\notin\m{P}'\text{ and }\exists(p,p')\in\m{E},p'\in\m{P}'\}$.
Then we can construct two execution states $\m{P}_1=\{p\mid p\in\m{P}'\text{ or }\exists p'\in\m{P}_1,(p,p')\in\m{E}\}$ and $\m{P}_2=\{p\mid p\in\m{S}\text{ or }\exists p'\in\m{P}_2,(p,p')\in\m{E}\}$.
From the convexity of $\m{P}'$ we can get $\m{P}'=\m{P}_1\setminus\m{P}_2$.
\end{proof}

Intuitively, the number of execution states increases linearly with the depth (i.e., the number of operators on the critical path) and exponential with the width (i.e., the number of concurrent operators) of a primitive graph.
While the number of execution states for a primitive graph can increase exponentially with the number of primitives, we observe that modern DNN architectures are generally {\em deep} instead of wide, resulting in a limited number of execution states and making it tractable to explicitly consider {\em all} execution states. 
For a primitive graph, we expect approximately $O(|\m{P}|)$ execution states and $O(|\m{P}|^2)$ candidate kernels for simplicity, where $|\m{P}|$ is the number of primitives in the graph.
Such simplification has also been adopted in prior work~\cite{tarnawski2021piper}.

Next, we describe \sys's {\em kernel identifier}, which uses a depth-first search algorithm for identifying {\em all} valid kernels in a primitive graph.

\begin{algorithm}[t]
\caption{Depth-first-search to identify kernels. \textproc{Profiling} takes a set of primitives as an input, generates a kernel for the defined tensor algebra computation, and profiles the runtime performance of the kernel. \textproc{Profiling} returns $\infty$ if such a kernel cannot be generated. We describe the kernel profiler in \Cref{subsec:profiler}. }
\label{alg1}
\small
\begin{algorithmic}[1]
\State {\bf Input:} A primitive graph $\m{G} = (\m{P}, \m{E})$
\State {\bf Output:} All possible kernels of $\m{G}$
\State $\m{B} = \{\}$ \Comment{\em $\m{B}$ is a database of $\m{G}$'s execution states}
\Function{Dfs}{$\m{X}$} \Comment{{\em $\m{X}$ is the current execution state}}
\For{$v\in\m{P} \setminus \m{X}$} \Comment{find a new primitive to execute}
\If{$\forall (u,v)\in\m{E}.u\in\m{X}$}
\State $\m{X}' = \m{X} \cup \{v\}$
\If{$\m{X}' \notin \m{B}$}
\State $\m{B} = \m{B} \cup \{\m{X}'\}$
\State \Call{Dfs}{$\m{X}'$}
\EndIf
\EndIf
\EndFor
\EndFunction
\State \Call{Dfs}{$\emptyset$}
\State $\m{K} = \{\}$ \Comment{\em $\m{K}$ is a database of $\m{G}$'s kernels}
\For{$\m{D}_1 \in \m{B}$}
\For{$\m{D}_2 \in \m{B}$}
\If{$\m{D}_1 \subset \m{D}_2$}
\State $\m{P}' = \m{D}_2 \setminus \m{D}_1$
\For{$\m{O} \subset \m{P}'$} \Comment{\em $\m{O}$ is a possible output set of $\m{P}'$}
\State $K, \er{latency} = \Call{Profiling}{\m{P}', \m{O}}$
\If{$\er{latency} \neq \infty$}
\State $\m{K} = \m{K} \cup \{K, \m{P}', \m{O}, \er{latency}\}$
\EndIf
\EndFor
\EndIf
\EndFor
\EndFor
\State {\bf Return} $\m{K}$
\end{algorithmic}
\end{algorithm}

\subsection{Kernel Identifier}
\label{subsec:kernel_identifier}
\Cref{alg1} shows an algorithm to identify all possible kernels in a primitive graph $\m{G}$.
The algorithm starts from an empty execution state $\m{X} = \emptyset$ and uses {\em depth-first} search (DFS) to enumerate all possible execution states, which are maintained in a database $\m{B}$.
For each pair of execution states $(\m{D}_1, \m{D}_2)$, their set difference identifies a set of primitives $\m{P}'= \m{D}_2 \setminus \m{D}_1$ that can potentially form a kernel.

Considering a subgraph $\m{G'=(P',E')}$, we now identify the inputs and outputs of $\m{G'}$.
Obviously, the inputs of $\m{G'}$ should be all the nodes of $\m{G'}$ with zero in-degree.
However, the outputs of $\m{G'}$ are not simply all the nodes with zero out-degree, since some nodes with positive out-degree may become the inputs for another candidate kernel subgraph.
We define the \emph{possible output set} of $\m{G'}$ as follows:

\begin{definition} [Possible output set]
    A set $\m{O}$ of nodes is a possible output set of a candidate subgraph $\m{G'=(P',E')}$ if and only if $\forall u\in\m{O}$, $\exists (u,v)\in\m{E}$ such that $v\in\m{P}\setminus\m{P'}$.
\end{definition}

Both $\m{P}'$ and $\m{O}$ are sent to the {\em kernel profiler} to generate a kernel for its primitives. If such a kernel can be successfully generated, the kernel profiler also returns the measured execution latency of the kernel; otherwise, the profiler returns infinity to indicate that $\m{P}'$ is not supported by \sys's DNN backends.
\Cref{subsec:profiler} introduces \sys's kernel profiler for generating and profiling kernels for a given set of primitives.
For the primitive graph in \Cref{fig:primitive_graph_kernel_mapping}, \Cref{fig:kernel_table} shows all valid kernels identified by \sys's kernel profiler.

\subsection{Kernel Orchestration Optimizer}
\begin{figure}
    \centering
    \includegraphics[width=\linewidth]{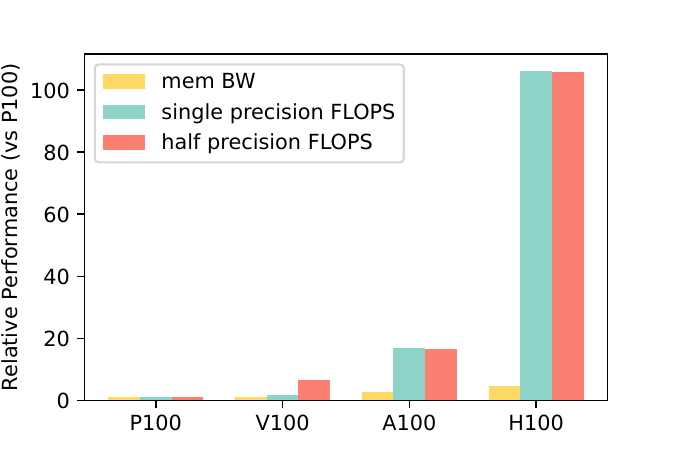}
    \caption{Comparing memory bandwidth and floating-point throughput across GPU generations. The y-axis is normalized to the performance of P100 GPUs.}
    \label{fig:memory_vs_compute}
\end{figure}

This section first defines the kernel orchestration problem and then formalizes it as a constrained optimization task. \sys uses a binary integer linear programming algorithm to discover a kernel orchestration strategy that minimizes the total execution cost on GPUs.

\paragraph{Kernel orchestration problem.} For a given primitive graph $\m{G}$, let $K_1, ..., K_M$ denote all the kernels discovered by the kernel identifier (see \Cref{subsec:kernel_identifier}).
The {\em kernel orchestration optimizer} takes a primitive graph $\m{G}$ and all identified kernels as input and discovers an {\em optimal} kernel execution strategy, which is represented by an $M$-dimensional binary vector $\vec{u} = \langle u_1,...,u_M\rangle$, where $u_i = 1$ if $K_i$ is executed in this strategy, and $u_i = 0$ otherwise. \sys follows prior work~\cite{jia2019taso} and assumes that the run time of an execution strategy is the summation of individual kernels' run times:
\begin{equation}
\label{eqn:objective}
\er{Cost}(\vec{u}) = \sum_{i=1}^{M} c_i u_i,
\end{equation}
where $c_i$ is the measured run time of kernel $K_i$. 

To discover a kernel execution strategy that minimizes run time, our key idea is to formalize kernel orchestration as a {\em binary linear programming} (BLP) task and find an optimal strategy using an off-the-shelf BLP solver, which will be introduced in \Cref{sec:impl}.
\sys uses two matrices $I$ and $O$ to represent the correlation between kernels and primitives. Specifically, $I$ and $O$ are both binary matrices of size $M \times |\m{P}|$, where $I_{ij} = 1$ if and only if primitive $p_j$ is an input of kernel $K_i$, and $O_{ij} = 1$ if and only if primitive $p_j$ is an input of kernel $K_i$.
Therefore, $\sum_{i=1}^{M} O_{ij} u_i$ calculates how many times primitive $p_j$ is executed.
A key difference between \sys and prior work is that existing tensor program optimizers partition computations defined in a DNN architecture into kernels in a {\em disjoint} way (i.e., no redundant computation), while \sys allows a primitive to be executed multiple times (i.e., $\sum_{i=1}^{M} O_{ij} u_i > 1$). 
This relaxation is based on an important observation that the floating-point throughput of modern GPUs is increasing at a much higher rate than their memory bandwidth, as shown in \Cref{fig:memory_vs_compute}.
Therefore, \sys allows primitives to be executed multiple times to opportunistically reduce memory accesses and kernel launch overheads. For example, \Cref{fig:kernel_mapping_strategy} shows an optimal kernel orchestration strategy that executes $p_1$ three times to reduce kernel launch and data movement overheads.

\paragraph{Comparison with tensor rematerialization.}
{\em Tensor rematerialization} is a common technique used in DNN training to reduce memory overhead of DNN training by discarding some intermediate results during forward processing and rematerializing the discarded tensors during backpropagation.
Tensor rematerialization reduces the peak memory usage of DNN training at the cost of re-executing some kernels multiple times in a training iteration.
Instead of reducing overall accesses to GPU device memory, tensor rematerialization actually increases memory accesses since some kernels are executed multiple times. 
In contrast, the primitive redundancy in \sys is designed to allow opportunistic fusion of kernels with overlapping primitives.

We now describe our binary linear programming formulation. For a given primitive graph $\m{G} = (\m{P}, \m{E})$, let $\m{T}\subseteq \m{P}$ denote the set of output primitives. The BLP task minimizes the objective defined in \Cref{eqn:objective} by considering the following constraints.

\paragraph{Output constraints.} The output constraints guarantee that a valid strategy executes all primitives that are the output of the DNN model:

\begin{equation}
    \sum_{i=1}^{M} O_{ij} u_i \geq 1 \quad \quad \forall p_j \in \m{T}. 
\end{equation}

\paragraph{Dependency constraints.} The dependency constraints check data dependencies between different kernels --- a kernel can be executed only if all its input tensors have been computed by prior kernels:

\begin{equation}
    \sum_{i=1}^{M} O_{ij} u_i \geq I_{k,j} u_k \quad \quad \forall p_j \in \m{P}, k\in [1,M]
\end{equation}

Note that \sys does not require all primitives to be executed, and our formulation implicitly performs pruning optimizations. %
\sys's kernel execution strategy can be seen as covering the whole computation graph with candidate kernel subgraphs, only with the constraints of subgraph dependencies.

%% file: implementation.tex
\section{Implementation}
\label{sec:impl}

This section presents our implementation of \sys, with a specific focus on the design of our operator fission engine and kernel orchestration optimizer.

\subsection{Operator Fission}
Our implementation represents a primitive graph in the ONNX 
 format~\cite{onnx}, which is widely used and mathematically complete.
In particular, ONNX operators \cite{onnx_operators} include all primitive types introduced in \Cref{sec:fission}\footnote{The {\tt Broadcast} primitive is performed implicitly during the computation of ONNX operators.}.
\sys's operator fission engine takes an ONNX graph as input and performs rule-based operator fission to construct an equivalent primitive graph. The output of the operator fission engine is also in the ONNX format.

\subsection{Kernel Orchestration Optimizer}
\label{subsec:profiler}

\paragraph{Kernel profiler}
\sys's {\em kernel profiler} takes a primitive graph as input, generates a GPU kernel for the input graph, and profiles the runtime performance of the kernel.
\sys follows prior work and requires that each primitive graph has exactly one output~\cite{tvm}.
As a result, each candidate kernel has an arbitrary number of input tensors and produces one output tensor.
For each candidate kernel, the profiler examines whether the subgraph includes any \textit{linear transformation primitive}.
If no linear transformation primitive is found, the profiler identifies the current candidate kernel as \textit{memory-intensive} and sends the subgraph to TVM's MetaSchedule for hyper-parameter tuning. Otherwise, the profiler identifies the current candidate kernel as \textit{compute-intensive}.

MetaScheduler~\cite{metascheduler} is a probabilistic scheduler DSL developed in TensorIR that unifies the pre-existing approaches (i.e., AutoTVM~\cite{tvm_auto_tuner} and Ansor~\cite{ansor}). \sys uses MetaScheduler to generate high-performance kernels for a given primitive graph and target hardware. Specifically, \sys lowers an ONNX graph to a TensorIR schedule given pre-defined scheduling rules and leverages the Ansor backend to generate the best performing schedule. 
The kernel profiler then measures the performance of the generated kernel and returns the measured latency, the lowered scheduler, and generated CUDA code back to the kernel orchestration optimizer.
Since memory-intensive kernels have relatively simple schedules, most of them can be tuned within 2 minutes through MetaScheduler.

For compute-intensive operators, \sys follows the heuristics used by prior work (e.g., EinNet~\cite{zheng2023einnet}) and directly lowers these primitive graphs to vendor libraries (e.g., cuDNN and cuBLAS) instead of TVM. This design is based on an observation that compute-intensive operators have more complicated schedules and are difficult to automatically generate a kernel that outperforms vendor-provided implementations within a short tuning time.
In the rare case where a compute-intensive primitive graph cannot match the parameters defined in vendor libraries, \sys rejects the candidate.
Our implementation uses cuDNN \cite{cudnn}, cuBLAS \cite{cublas} and TensorRT \cite{tensorrt} as potential backends for compute-intensive operators.

\sys formalizes kernel orchestration as a binary linear programming (BLP) problem by constructing the constraints identified in \Cref{sec:mapping}, and solves the BLP using PuLP \cite{mitchell2011pulp}, a Python linear programming solver.
The biggest subgraph in our test cases has 584 execution states and 3078 candidate kernels.
PuLP can converge to an optimal solution within 1000 seconds for all primitive graphs in our evaluation.

\begin{figure*}
    \centering
    \includegraphics[width=\linewidth,keepaspectratio]{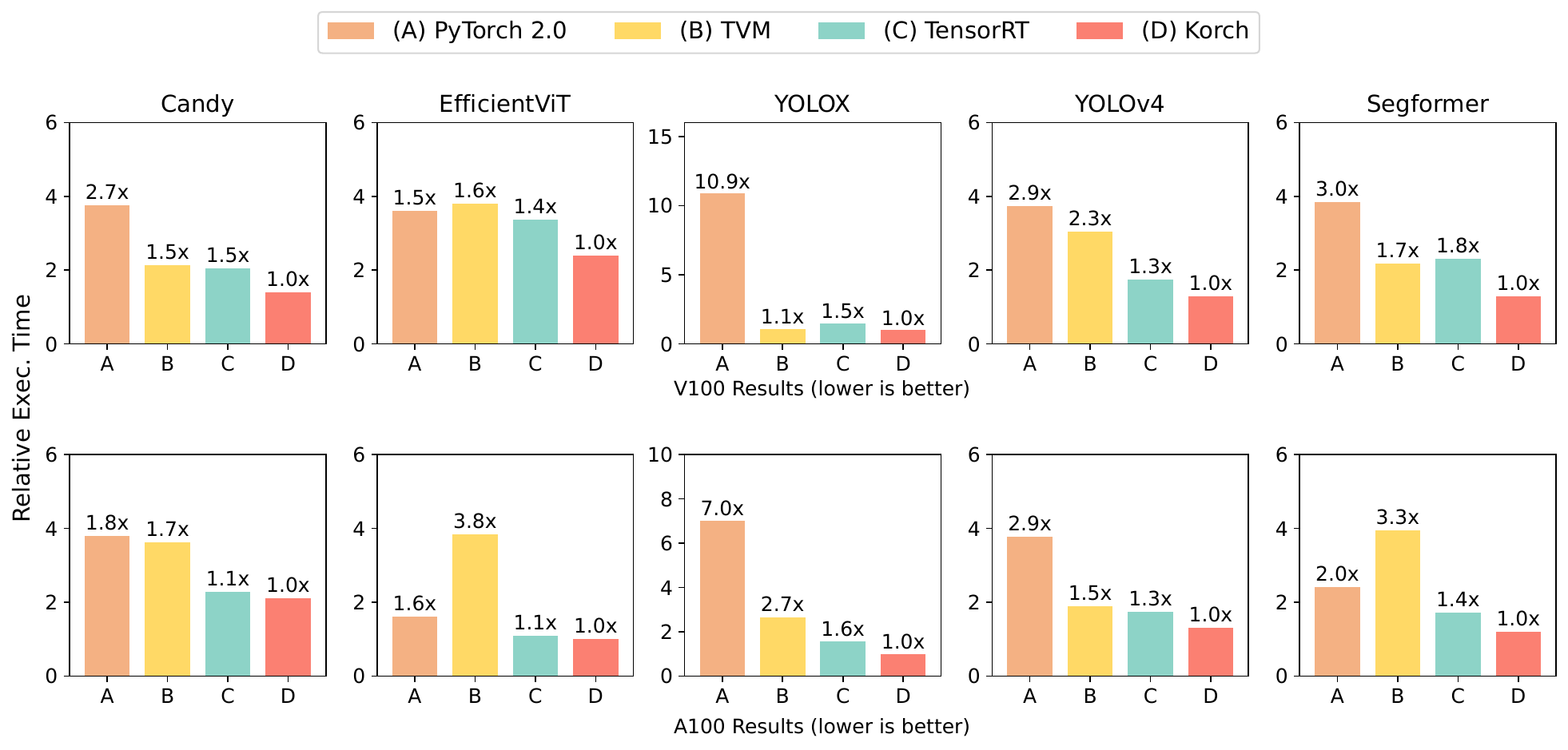}
    \caption{End-to-end performance comparison on V100 and A100.}
    \label{fig:eval-e2e}
\end{figure*}
\subsection{Executable Generator}
After generating an optimal kernel orchestration strategy and the CUDA code for all selected candidate kernels, \sys stitches the kernels' implementations together by respecting data dependencies between these kernels.
As a result, the end-to-end execution latency of the entire computation graph is consistent with the objective function of the ILP task (i.e., \Cref{eqn:objective}).
Note that \sys only considers sequential execution of the orchestrated kernels and does not consider inter-kernel optimizations such as CUDA multi-streaming.
By modifying the ILP problem formulation, it is possible to take into account more advanced kernel execution strategies, which we leave as future work.

%% file: evaluation.tex
\section{Evaluation}

\subsection{Experimental Setup}

\paragraph{Platform}
Our evaluation is performed on an Nvidia V100 GPU and an A100 GPU.
For the V100, we use an AWS\linebreak p3.8xlarge instance \cite{p3.2xlarge}, 
which is equipped with a 32-core Intel Xeon CPU, four V100 16GB SXM GPUs, CUDA 11.7 and runs Ubuntu 20.04.
For the A100, we use the Perlmutter supercomputer, which runs on SUSE Linux Enterprise Server 15 SP4 OS and is equipped with with a 64-core AMD EPYC CPU, four A100 80GB SXM GPUs and CUDA 11.7 on each node.

\paragraph{Workloads}
We use five real-world DNNs to evaluate \sys.
Candy \cite{johnson2016perceptual} is a convolutional neural network (CNN) for fast style transfer.
YOLOv4 \cite{bochkovskiy2020yolov4} and YOLOX-Nano \cite{ge2021yolox} are CNN-based object detectors.
Segformer \cite{xie2021segformer} is a vision Transformer for semantic segmentation.
EfficientViT \cite{cai2022efficientvit} is a vision Transformer backbone for high-resolution low-computation vision tasks.
For all those workloads, we evaluate with a batch size of one.
The input resolutions for Segformer and EfficientViT are 512$\times$512 and 2048$\times$2048, respectively.
For other CNNs, we use their default input resolutions: 224$\times$224 for Candy and 416$\times$416 for YOLOs.
On V100 GPUs, all DNNs are executed in 32-bit floating points (FP32).
On A100 GPUs, we enable tensor cores and execute the DNNs using TF32.

\subsection{End-to-end Performance}
We first evaluate the end-to-end inference latency on a single GPU, and compare \sys with PyTorch 2.0.1, TVM 0.11~\cite{tvm} (commit hash 61a4f21) and TensorRT 8.2~\cite{tensorrt}.

\Cref{fig:eval-e2e} shows the end-to-end results on V100 and A100. \sys achieves up to 1.7$\times$ speedup on V100 and up to 1.6$\times$ speedup on A100.
The average speedups on V100 and A100 are 1.39$\times$ and 1.30$\times$, respectively.
Among the five DNN workloads, Segformer and EfficientViT are vision Transformers and heavily optimized by existing DNN frameworks.
With the optimal kernel orchestration strategy, \sys can still reduce their latency by up to 1.7$\times$ and 1.4$\times$.
For conventional CNNs such as Candy, YOLOv4 and YOLOX-Nano, \sys outperforms existing frameworks by 1.31$\times$ on average.
We analyze the optimization improvement and show the new orchestration strategies discovered by \sys in \Cref{subsec:eval_tensorrt} and \Cref{subsec:case_study}.

We observe that \sys achieves a more significant performance improvement on V100 than on A100.
As verified by prior approach~\cite{zheng2023einnet}, advanced GPUs like A100 should get more performance benefits theoretically from tensor program optimizations (e.g., kernel fusion), since they offer a higher ratio between computation capability and memory bandwidth (see \Cref{fig:memory_vs_compute}).
The reverse behaviors in our experiments can be attributed to two reasons.
First, kernel orchestration not only optimizes memory-intensive operators, which are the major source of acceleration from prior kernel fusion approaches, but also improves compute-intensive operators' performance.
We further discuss the difference in \Cref{subsec:case_study} and provide a case study (i.e., \Cref{fig:qkv_case}) to show how \sys considers different data layouts for those compute-intensive operators.
Second, \sys's kernel orchestration relies on TVM for flexible code generation for candidate kernels.
Our evaluation also shows that TVM performs worse than highly-optimized TensorRT on A100, indicating that there still remains potential space for improving \sys by leveraging more advanced kernel generation techniques.

\subsection{Adaptation Study over TensorRT}
\label{subsec:eval_tensorrt}

\begin{figure}
    \centering
    \includegraphics[width=0.45\linewidth]{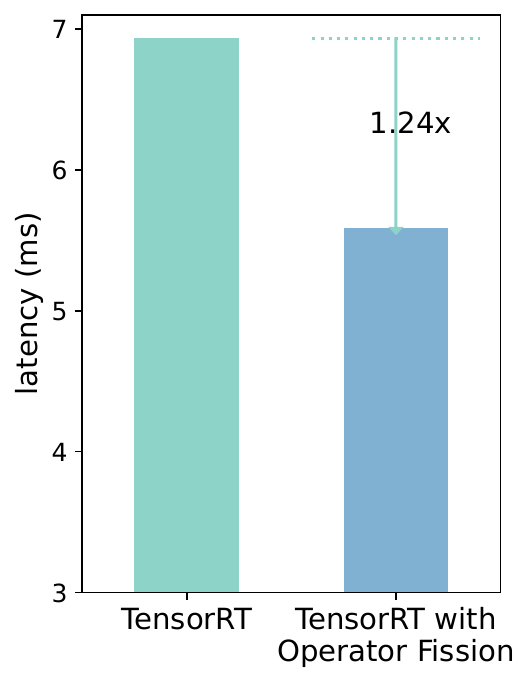}
    \caption{speedup of operator fission: feeding the primitive graph instead of computation graph to TensorRT leads to 1.24x speedup.}
    \label{fig:ablation}
\end{figure}

In this section, we transplant the implementation of our proposed operator fission to TensorRT to show its effectiveness.
Specifically, instead of using the ILP-based kernel orchestration in \Cref{fig:kernel_mapping}, we directly feed the post-operator-fission primitive graph to TensorRT, which then decides the kernel orchestration strategy and uses its own kernels for code generation.
Since TensorRT is heavily optimized for the Transformer models, we select Segformer for this study to make the comparison more persuasive.
\Cref{fig:ablation} shows the performance comparison on Segformer and we can see that merely applying operator fission leads to a 1.24$\times$ speedup on V100, compared with directly using TensorRT.

\subsection{Case Study}
\label{subsec:case_study}

To understand how \sys's optimal kernel orchestration technique enlarges the optimization space, we study some orchestration strategies found by \sys in detail.
All results in this section are measured on Nvidia V100 GPU.

\begin{figure}
    \centering
    \subfloat[TensorRT's strategy.]{
        \includegraphics[scale=1]{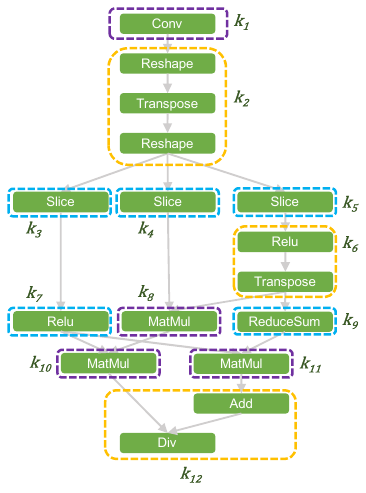}
        \label{fig:qkv_tensorrt}
    }
    
    \subfloat[\sys's strategy.]{
        \includegraphics[scale=1]{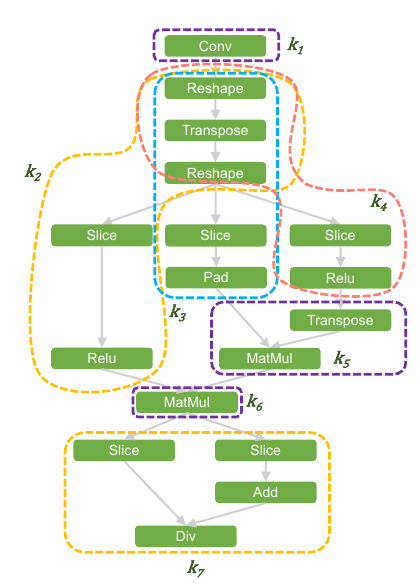}
        \label{fig:qkv_ours}
    }
    \caption{Comparison of kernel orchestration strategy on an EfficientViT attention block. Before kernel orchestration, \sys first applies graph transformation in \Cref{fig:qkv_taso}.}
    \label{fig:qkv_case}
\end{figure}

\begin{figure*}
    \centering
    \subfloat[Merge the blue \texttt{MatMul} and \texttt{ReduceSum}.]{
        \includegraphics[width=0.32\textwidth]{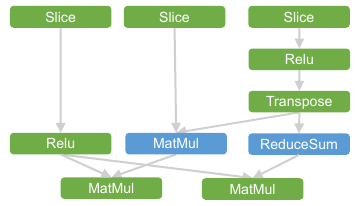}
    }
    \subfloat[Merge the orange \texttt{MatMuls}.]{
        \includegraphics[width=0.32\textwidth]{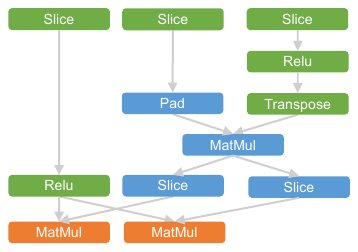}
    }
    \subfloat[After transformation.]{
        \includegraphics[width=0.32\textwidth]{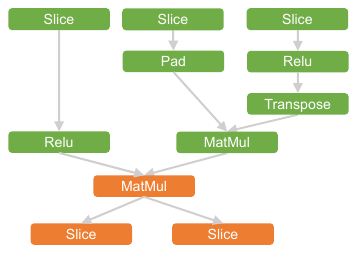}
    }
    \caption{Primitive graph transformation from \Cref{fig:qkv_tensorrt} to \Cref{fig:qkv_ours}. Nodes with same color indicate a transformation on the primitive graph. Merging of \texttt{ReduceSum} and \texttt{MatMul} is done similarly with \Cref{fig:fission}.}
    \label{fig:qkv_taso}
\end{figure*}

\begin{figure}
    \centering
    \includegraphics[width=0.5\linewidth]{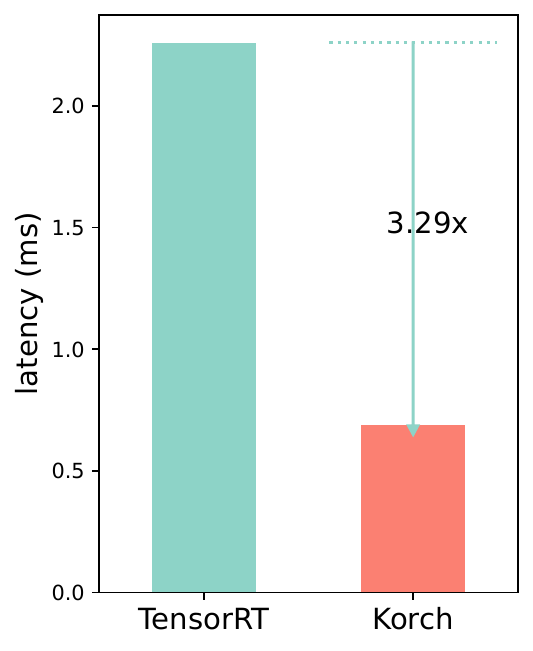}
    \caption{Case study of EfficientViT attention block: Korch's strategy (\Cref{fig:qkv_ours}) achieves 3.29x speedup over TensorRT's strategy (\Cref{fig:qkv_tensorrt}).}
    \label{fig:case_vit}
\end{figure}

\paragraph{Redundant Computing.}
\Cref{fig:qkv_case} compares the kernel orchestration strategy of TensorRT and \sys in the EfficientViT attention block.
\sys first applies transformations on the primitive graph to merge \texttt{ReduceSum} and \texttt{MatMul}s, generating the graph from \Cref{fig:qkv_taso} to \Cref{fig:qkv_ours}.
During kernel orchestration, \sys fuses consecutive memory-intensive operators (e.g., $k_2$ in \Cref{fig:qkv_ours}) and opportunistically fuses \texttt{Transpose} with \texttt{Matmul} to optimize data layout (e.g., comparing $k_8$ in \Cref{fig:qkv_tensorrt} and $k_5$ in \Cref{fig:qkv_ours}), which is not considered by TensorRT.
In particular, the input matrix to $k_8$ in \Cref{fig:qkv_tensorrt} has an extreme aspect ratio of 1024:1.
After changing the data layout, $k_5$ in \Cref{fig:qkv_ours} is 3.52x faster than $k_8$ in \Cref{fig:qkv_tensorrt}, with the same TensorRT MatrixMultiply backend.
Moreover, \sys allows the \texttt{Reshape}+\texttt{Transpose}+\texttt{Reshape} operators to be executed multiple times in three kernels (i.e., $k_2$, $k_3$, and $k_4$ in \Cref{fig:qkv_ours}).
Since all these operators are memory-intensive, their latency is mainly determined by the memory I/O overhead.
\sys's mapping strategy reduces the total number of kernels and therefore improves the overall latency.
\sys only uses 7 kernels for the EfficientViT attention block, which saves 5 kernels and achieves 3.29$\times$ speedup over TensorRT for the entire subgraph, as shown in \Cref{fig:case_vit}.
If we directly apply kernel orchestration in \Cref{fig:qkv_tensorrt}'s graph without primitive graph transformations, the optimizer also needs to do redundant computation (i.e., moving \texttt{Transpose} in $k_6$ into $k_8$ and $k_9$) to optimize the data layout of \texttt{MatMul} in $k_8$.
However, redundant computing is not considered in prior work, so this data layout transformation is out of the optimization space of prior work.

\begin{figure*}
    \centering
    \subfloat[TVM's kernel orchestration strategy.]{
        \includegraphics[width=0.48\textwidth]{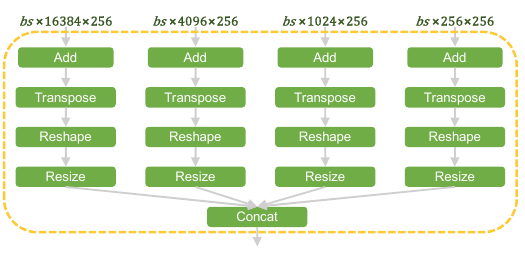}
        \label{fig:fusion_tvm}
    }
    \subfloat[\sys's kernel orchestration strategy with batch size 16.]{
        \includegraphics[width=0.48\textwidth]{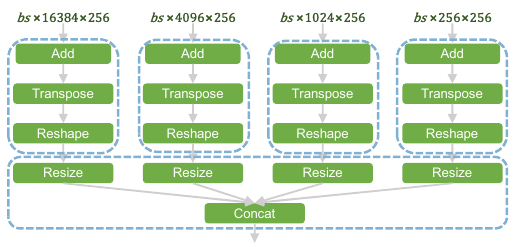}
        \label{fig:fusion_ours}
    }
    \caption{Two different kernel orchestration strategy on a Segformer subgraph. TVM will always choose strategy (a), but \sys will choose strategy (a) with batch size 1 and strategy (b) with batch size 16.}
    \label{fig:greedy_fusion_case}
\end{figure*}

\begin{figure}
    \centering
    \subfloat[TensorRT: 0.0911ms.]{
        \includegraphics[scale=0.85]{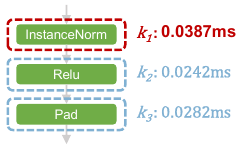}
        \label{fig:norm_tensorrt}
    }
    \subfloat[\sys: 0.0692ms.]{
        \includegraphics[scale=0.9]{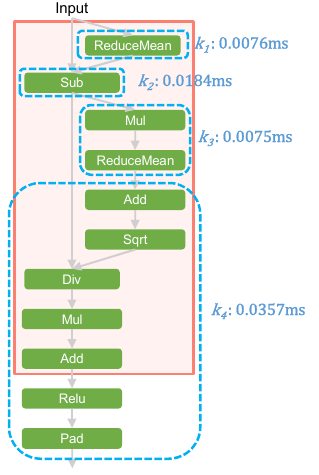}
        \label{fig:norm_ours}
    }
    \caption{Comparison of kernel orchestration strategy and kernel latency on a common subgraph pattern in Candy. The red frame in \Cref{fig:norm_ours} is the opeartor fission of \texttt{InstanceNorm}. \sys outperforms TensorRT by 1.32x on this subgraph.}
    \label{fig:norm_case}
\end{figure}

\paragraph{Map one operator to different kernels.}
To optimize Segformer, \sys uses the transformations demonstrated in \Cref{fig:primitive_graph} to decompose {\tt Softmax}, and then applies similar kernel mapping as shown in \Cref{fig:kernel_mapping_strategy}.
With this strategy, \sys maps {\tt Softmax} to all four kernels in \Cref{fig:primitive_graph} and outperforms TensorRT by 1.50$\times$ for this self-attention block.
Moreover, to optimize CNNs such as Candy, \sys can break the boundary of operators during fusion.
\Cref{fig:norm_tensorrt} shows a common network pattern in Candy. As can be seen in \Cref{fig:norm_case}, TensorRT maps {\tt InstanceNorm}, {\tt ReLU} and {\tt Pad} to individual kernels, but \sys can decompose {\tt InstanceNorm} and fuse part of it into the subsequent {\tt ReLU} and {\tt Pad} operators, achieving 1.32$\times$ speedup for this graph pattern.

\begin{figure}
    \centering
    \includegraphics[width=\linewidth]{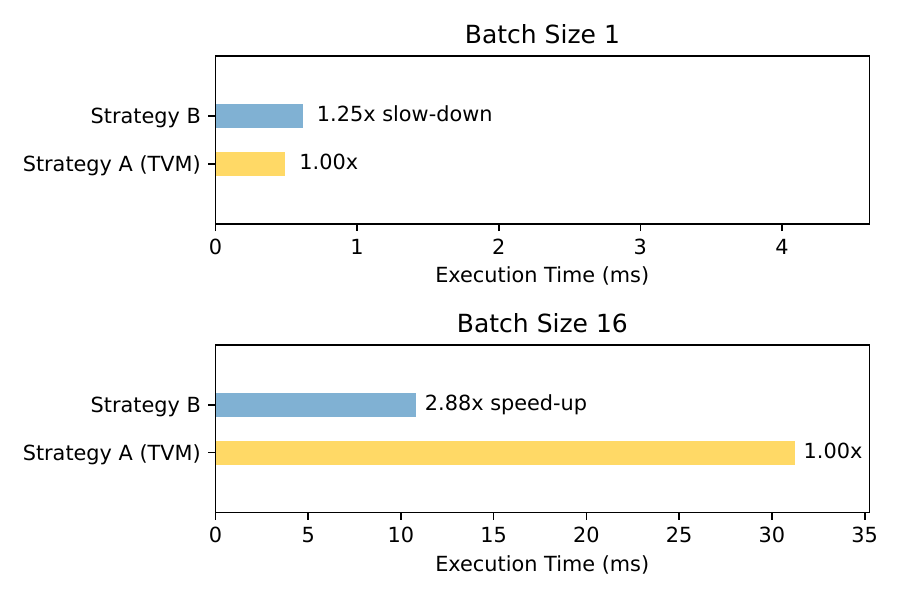}
    \caption{Comparison of different kernel orchestration strategies on a Segformer subgraph. TVM will always choose strategy A, but \sys will choose strategy A for batch size 1 and strategy B for batch size 16.}
    \label{fig:greedy}
\end{figure}

\paragraph{Greedy fusion can be suboptimal.}
\Cref{fig:greedy_fusion_case} shows two different kernel orchestration strategies for a subgraph of Segformer. 
Since all operators are memory-intensive and fusable, TVM directly fuses the entire subgraph into one kernel.
However, as shown in \Cref{fig:greedy}, fusing the entire graph is efficient with a batch size of 1 but suboptimal with a batch size of 16, since TVM cannot generate a highly optimized kernel for the entire graph.
Without modifying the TVM code generation backend, using multiple kernels (\Cref{fig:fusion_ours}) to execute this graph leads to 2.24$\times$ speedup.

\subsection{Tuning Time}

\Cref{tab:tuning_time} shows the number of primitive graph nodes, candidate kernels, and overall tuning time on Perlmutter for all DNNs in our evaluation.
Theoretically the number of candidate kernels should be quadratic with the number of nodes, but we observe that the actual number of candidate kernels is far smaller than that.
In our implementation, most of the candidate kernels can be rejected with simple heuristics, such as too many operators to generate within one kernel or including multiple linear transformation primitives.

Most of \sys's tuning time is consumed by TVM's Meta\-Scheduler for memory-intensive kernels.
We utilize the TVM database to avoid tuning the same candidate kernel multiple times.
Besides, the tuning of candidate kernels can be parallelized across multiple GPUs.
The longest tuning time is 12.2 hours, in which case a single kernel takes 12 hours to tune by the TVM MetaScheduler.
Given that kernel tuning is one-time cost, \sys's tuning overhead is acceptable in real world applications.

\begin{table}
\caption{Number of primitive graph nodes, candidate kernels and end-to-end tuning time on Perlmutter for all the evaluated models.}
\label{tab:tuning_time}
\begin{tabular}{lrrrr}
\toprule
Model        & \multicolumn{1}{r}{\# Nodes} & \makecell{\# Candidate\\ Kernels} & \multicolumn{1}{r}{Tuning Time} \\ \hline
Candy        & 184                         & 1031                                    & 5.5h                                 \\
EfficientViT & 380                         & 2174                                    & 11.5h                                \\
YOLOX        & 367                         & 3361                                    & 2.8h                                 \\
YOLOv4       & 569                         & 4644                                    & 12.2h                                \\
Segformer    & 672                         & 11400                                   & 9.2h                                 \\ \bottomrule
\end{tabular}
\end{table}

%% file: related.tex
\section{Related Work}

\paragraph{Kernel orchestration.} Existing deep learning frameworks use rule-based operator fusion strategies to map tensor operators to kernels on modern hardware platforms. For example, DNNFusion~\cite{dnnfusion} starts fusion at the one-to-one operator with the minimum intermediate result, and then tries to fuse its successors and predecessors with a greedy strategy.
A key difference between \sys and these frameworks is that \sys provides a new and systematic approach to optimizing kernel orchestration by first applying operator fission to decompose tensor operators into basic primitives and then using binary programming to discover an optimal kernel execution strategy for each graph of primitives.
PyTorch 2.0 \cite{pytorch} will also decompose operators to primitives, but this is only for reducing the number of operators from $\sim 2000$ to a stable set of $\sim 250$, in order to simplify engineering effort on new operators and new hardwares.

\paragraph{Dynamic programming solutions.}
Dynamic programming (DP) is widely used for fusion in domain-specific compilers.
For example, PolyMageDP~\cite{jangda2018effective} uses a DP fusion algorithm for image processing pipelines. However, there are two key differences between \sys and this kind of work.
First, DP solutions directly fuse operators and thus miss cross-operator fusion opportunities, while \sys applies operator fission to decompose operators into fine-grained primitives to enable a diversity of cross-operator fusion optimizations for maximizing kernel efficiency.
Second, \sys considers re-executing primitives to enable additional kernel fusion opportunities.
Re-execution of operators is challenging to consider with DP but is naturally modeled by \sys's binary linear programming.

\paragraph{Graph optimization.}
TensorFlow~\cite{abadi2016tensorflow}, TensorRT~\cite{tensorrt}, and MetaFlow~\cite{MetaFlow} optimize operator graphs by applying graph transformation rules designed by domain experts. TASO~\cite{jia2019taso} automatically generates and verifies graph transformations for a given set of tensor operators, and uses backtracking search to apply the generated transformations. PET~\cite{wang2021pet} extends TASO by generating partially equivalent transformations and their correction kernels, thus enabling a larger search space for graph optimization. \sys's kernel orchestration optimizations are orthogonal and can be combined with existing graph-level optimizations. In particular, \sys uses the graph transformations discovered by TASO to optimize primitive graphs (see \Cref{sec:fission}).

\paragraph{Hand-optimized kernels.}
Hand-optimized kernels written by human experts can achieve near-optimal performance for dedicated subgraphs.
For example, FlashAttention~\cite{dao2022flashattention} optimizes device memory access \emph{within} a kernel for attention algorithm.
This kind of work is orthogonal to \sys, which optimizes the orchestration of DNN operators \emph{across} multiple kernels.
In fact, hand-optimized kernels can be directly incorporated into \sys as an additional backend of the kernel profiler.
For example, FlashAttention can be integrated in \sys for generating attention kernels, which will be considered by \sys’s kernel orchestration optimizer when it identifies an attention subgraph.

\paragraph{Kernel generation.}
Recent work has proposed a variety of approaches to automatically generating hardware-specific kernels for DNN computation. For example, Halide~\cite{halide, autohalide} decouples an image processing program into an algorithm and a schedule, and proposes several strategies to discover highly optimized schedules. This idea of separating algorithm with schedule has been adopted by many deep learning compilers~\cite{flextensor, tvm, adams2019learning, anderson2021efficient}. In particular, TVM~\cite{tvm_auto_tuner} uses a learning-based approach to predicting the cost of a schedule and discovering efficient schedules in a pre-defined schedule space. Ansor~\cite{ansor} automatically generates schedule templates, resulting in more performant schedules than TVM. \sys directly uses TVM and Ansor for generating high-performance kernels for a given set of primitives.
There're also some emerging kernel generation frameworks such as Triton \cite{tillet2019triton} and Hidet \cite{ding2023hidet}. All of this kind of work can be added to \sys's backend as a choice for kernel profiling.

%% file: future_work.tex
\section{Future Work}

\paragraph{Tuning time acceleration.}

The majority of \sys's end-to-end tuning time is consumed in candidate kernel profiling.
Although the overall tuning time on evaluated DNNs is acceptable, the number of candidate kernels considered by \sys is exponential in width, which limits \sys's extension to general tensor programs.
Building a lightweight cost model to quickly discard inefficient candidates is a possible solution.

\paragraph{Extending BLP problem formulation.}
The binary linear programming (BLP) formulation in this paper has some room for improvement.
Currently, \sys only considers candidate subgraphs with a single output.
If multiple outputs of a subgraph are properly defined, we can remove this single-output constraint to make the problem formulation more general.
Moreover, it is possible to take different data layouts into account in the BLP problem.
For each candidate kernel $K$, we can specify the data layout (e.g., \texttt{NCHW}, \texttt{NHWC}) of each input and output.
Then the BLP solver can automatically choose the optimal data layout during calculation of the computation graph.

%% file: conclusion.tex
\section{Conclusion}
This paper identifies the kernel orchestration task in tensor program compilation and presents \sys, a systematic approach to discovering optimal kernel orchestration strategies for tensor programs.
\sys applies operator fission to decompose tensor operators into fine-grained primitives, formalizes kernel orchestration as a binary linear programming task, and uses an off-the-shelf solver to generate an optimal strategy.
\sys outperforms existing tensor program optimizers by up to 1.7$\times$ on modern GPU architectures.

\section*{Acknowledgement}
We thank the anonymous reviewers and our shepherd Alex Reinking for their feedback on this work.
We thank Jiachen Yuan for open source of this project.
This research is partially supported by NSF awards (CNS-2147909, CNS-2211882, and CNS-2239351), NSFC for Distinguished Young Scholar\linebreak(62225206), National Natural Science Foundation of China (U20A20226), and research awards from Amazon, Cisco,\linebreak Google, Meta, Oracle, Qualcomm, and Samsung. 
The views and conclusions contained in this document are those of the authors and should not be interpreted as representing the official policies, either expressed or implied, of any sponsoring institution, the U.S. government or any other entity.

%% file: appendix.tex
\appendix
\section{Artifact Appendix}

\subsection{Abstract}

This artifact appendix helps the readers run the artifact and reproduce results of the paper: Optimal Kernel Orchestration for Tensor Programs with Korch.

\subsection{Artifact check-list (meta-information)}

{\small
\begin{itemize}
  \item {\bf Run-time environment: }AWS p3.2xlarge instance.
  \item {\bf Hardware: }Nvidia Tesla V100-SXM2-16GB.
  \item {\bf Metrics: }Execution time.
  \item {\bf Output: }Execution time.
  \item {\bf Experiments: }Adaption study of operator fission over TensorRT (\Cref{fig:ablation}), end-to-end evaluation (\Cref{fig:eval-e2e}) on two models.
  \item {\bf How much disk space required (approximately)?: }50GB.
  \item {\bf How much time is needed to prepare workflow (approximately)?: }1.5 hour.
  \item {\bf How much time is needed to complete experiments (approximately)?: }3 hours + 7 hours.
  \item {\bf Publicly available?: }Yes.
  \item {\bf Code licenses (if publicly available)?: }Apache License v2.0.
  \item {\bf Archived (provide DOI)?: }\url{https://doi.org/10.5281/zenodo.10839367}.
\end{itemize}
}

\subsection{Description}

\subsubsection{How to access}

The artifact is available on GitHub: \url{https://github.com/humuyan/ASPLOS24-Korch-AE} and Zenodo: \url{https://doi.org/10.5281/zenodo.10839367}, which includes Korch's source code, two benchmark models in ONNX format and documentation for installation and running.  

\subsubsection{Hardware and software dependencies}

This artifact is evaluated on AWS p3.2xlarge instance, which is equipped with an Intel(R) Xeon(R) CPU E5-2686 v4 2.30GHz CPU and an Nvidia Tesla V100-SXM2-16GB GPU.
AWS\linebreak p3.2xlarge instance runs Deep Learning AMI GPU PyTorch 1.13.0 (Ubuntu 20.04) environment.
The artifact relies on CUDA 11.7, cuDNN 8.6.0, TensorRT 8.2.0.6 and TVM 0.11 (commit hash 61a4f21).

\subsubsection{Models}

We evaluate Korch on two DNN models: Candy and Segformer.
The ONNX files of these two models can be accessed in the \texttt{cases/} directory of GitHub repository.
\texttt{candy.onnx} is from \url{https://github.com/onnx/models/tree/bec48b6a70e5e9042c0badbaafefe4454e072d08/validated/vision/style_transfer/fast_neural_style/model}, and \\\texttt{segformer.onnx} is exported from \url{https://huggingface.co/nvidia/segformer-b0-finetuned-ade-512-512}.

\subsection{Installation}

See Environment Preparation section in README.

\subsection{Experiment workflow}

See Run TensorRT Baseline section and Run Korch section in README.
The following experiments are included:
\begin{itemize}
    \item Adaption study of operator fission over TensorRT. This will take about 5 minutes.
    \item End-to-end evaluation of Korch on Candy model. This will take about 3 hours in the default configuration.
    \item End-to-end evaluation of Korch on Segformer model. This will take about 7 hours in the default configuration.
\end{itemize}

\subsection{Experiment customization}

The default configuration of this artifact does not enable TensorRT backend, since it will at least double the tuning time with only a marginal speed-up.
Although TensorRT can be enabled by changing \texttt{enable\_trt} to \texttt{True} on line 13 of \texttt{framework/calc.py}, we recommend users to use the default configuration to save the tuning time without harming the functional integrity test of Korch system.

\subsection{Notes}

The majority of Korch's end-to-end tuning time will be consumed in candidate kernel profiling.
There will be a \texttt{tqdm} progress bar showing ETA.
Korch uses TVM database to store profiled results during candidate kernel profiling, so users can halt the program and resume at any time during this progress.

\subsection{Evaluation and expected results}

For adaption study over TensorRT, the results should be similar with \Cref{fig:ablation}.
For end-to-end result on Candy and Segformer, the result of artifact's default configuration will be slightly slower than \Cref{fig:eval-e2e}, since TensorRT backend is disabled.